\documentclass[prd,preprint,superscriptaddress,showpacs,byrevtex]{revtex4}
\usepackage{bm}
\def\fsl#1{\setbox0=\hbox{$#1$}           % set a box for #1
   \dimen0=\wd0                                 % and get its size
   \setbox1=\hbox{/} \dimen1=\wd1               % get size of /
   \ifdim\dimen0>\dimen1                        % #1 is bigger
      \rlap{\hbox to \dimen0{\hfil/\hfil}}      % so center / in box
      #1                                        % and print #1
   \else                                        % / is bigger
      \rlap{\hbox to \dimen1{\hfil$#1$\hfil}}   % so center #1
      /                                         % and print /
   \fi}                                         %
\newcommand{\be}{\begin{equation}}
\newcommand{\ee}{\end{equation}}
\newcommand{\bea}{\begin{eqnarray}}
\newcommand{\eea}{\end{eqnarray}}
\newcommand{\beq}{\begin{equation}}
\newcommand{\eeq}{\end{equation}}
\newcommand{\beqs}{\begin{eqnarray}}
\newcommand{\eeqs}{\end{eqnarray}}

\begin{document}

\title{ Correct Definition of Color Singlet P-Wave Non-Perturbative Matrix Element of Heavy Quarkonium Production }
\author{Gouranga C Nayak }\thanks{G. C. Nayak was affiliated with C. N. Yang Institute for Theoretical Physics in 2004-2007.}
\affiliation{ C. N. Yang Institute for Theoretical Physics, Stony Brook University, Stony Brook NY, 11794-3840 USA}
%
%\date{\today}
\begin{abstract}
Recently we have proved factorization of infrared divergences in NRQCD S-wave
heavy quarkonium production at high energy colliders at all orders in coupling constant.
One of the problem which still exists in the higher order pQCD calculation of color singlet P-wave heavy
quarkonium production/anihillation is the appearance of non-canceling infrared divergences due to real soft gluons
exchange, although no such infrared divergences are present in the color singlet S-wave heavy quarkonium. In this paper
we find that since the non-perturbative matrix element of the color singlet P-wave
heavy quarkonium production contains derivative operators, the gauge links are necessary to make it gauge invariant and
be consistent with the factorization of such non-canceling infrared divergences at all orders in coupling constant.
\end{abstract}
\pacs{ 12.38.Lg; 12.38.Aw; 14.40.Pq; 12.39.St }
\maketitle
\pagestyle{plain}
\pagenumbering{arabic}
\section{Introduction}
Recently we have proved factorization of infrared divergences in NRQCD S-wave
heavy quarkonium production at high energy colliders at all orders in coupling constant in \cite{nknr}.
In this paper we extend this to prove factorization of infrared divergences in color singlet P-wave
heavy quarkonium production at high energy colliders at all orders in coupling constant. We predict the
correct definition of the color singlet P-wave non-perturbative matrix element of heavy quarkonium production.

Since the discovery of $j/\psi$ \cite{jpsi} there have been lot of experimental progress in the measurement of heavy
quarkonium production/decay both at fixed target experiments and at collider experiments. On the theoretical side
the understanding of heavy quarkonium production/decay proceeds in two steps, 1) the production/anihilation of heavy
quark-antiquark pair which can be calculated by using pQCD \cite{pqcd,pqcd1} and 2) the formation of heavy quarkonium from heavy
quark-antiquark pair which involves non-perturbative QCD which is not solved yet. Hence, unlike QED, the non-perturbative matrix
element of the heavy quarkonium in QCD is extracted from the experiment.

Assuming that the factorization theorem holds in QCD, the heavy quarkonium production from color singlet heavy quark-antiquark pair is given by
\bea
d\sigma_{pp \rightarrow H +X(P_T)} = \sum_{i,j}\int dx_1 dx_2 f_{i/p}(x_1,Q) f_{j/p}(x_2,Q) ~d{\hat \sigma}_{ij \rightarrow Q{\bar Q}[^{2S+1}L_J] +X(P_T)} ~<0|{\cal O}_H|0>
\label{cssip}
\eea
where the orbital angular momentum quantum number $L=0$ corresponds to S-wave heavy quarkonium such as $\eta_c,~j/\psi$
and $L=1$ corresponds to P-wave heavy quarkonium such as $h_c,~\chi_J$ where
$S$ and $J$ are the spin and total angular momentum quantum numbers of the heavy
quarkonium. The partonic level cross section for color singlet heavy quark-antiquark
($Q{\bar Q}$) pair production is given by $d{\hat \sigma}_{ij \rightarrow Q{\bar Q}[^{2S+1}L_J] +X(P_T)}$
which can be calculated by using pQCD where $i,j=q,{\bar q},g$. The $f_{i/p}(x,Q)$ is the parton
distribution function of the parton $i$ inside the hadron $p$.
The non-perturbative matrix element of the heavy quarkonium $(H)$
is given by $<0|{\cal O}_H|0>$. The heavy quarkonium is also denoted by $J^{PC}$ where
$P=(-1)^{L+1}$ and $C=(-1)^{L+S}$.

Since the partonic level cross section $d{\hat \sigma}_{ij \rightarrow Q{\bar Q}[^{2S+1}L_J] +X(P_T)}$
is calculated by using pQCD one expects to encounter non-canceling infrared divergences when
pQCD calculation is performed at higher orders in the QCD coupling constant. There are two types
of infrared divergence in this case, 1) infrared divergence due to the Coulomb gluon exchange and 2)
infrared divergence due to real soft gluon emission/absorption (eikonal gluon) \cite{cac}.

For S-wave and P-wave heavy quarkonium the appearance of infrared divergence due to Coulomb gluon exchange
(which is also known as $\frac{1}{v}$ divergence \cite{bar1,bar2,bar3,bar4} in the limit $v \rightarrow 0$ where $v$ is
the relative velocity between heavy quark and antiquark) is usually handled by
normalization of the bound state wave function similar to QED \cite{harris} where the Coulomb potential is used.

However, one may expect non-canceling infrared divergence due to emission/absorption of real soft gluon
(eikonal gluon). In the higher order pQCD calculation of
the color singlet S-wave heavy quark-antiquark production/anihillation one finds the absence of such infrared divergence
\cite{bar1,lep}. Hence the non-perturbative effects can be factored into the non-perturbative wave function at the origin
$R(0)$ for the S-wave heavy quarkonium. In terms of heavy quark fields one finds that the definition of the
gauge invariant non-relativistic wave function at the origin of the color singlet S-wave heavy quarkonium is given by \cite{nrqcd}
\bea
|R(0)|^2 =\frac{2\pi }{3}<0|{\cal O}_H|0> =\frac{2\pi }{3} <0|\chi^\dagger \xi a^\dagger_Ha_H \xi^\dagger \chi |0>
\label{sw}
\eea
where $\chi$ is the two component Dirac spinor field that creates a heavy quark
and $\xi$ is the two component Dirac spinor field that annihilates a heavy quark and
$a^\dagger_H$ is the creation operator of the heavy quarkonium $H$. In eq. (\ref{sw})
the non-perturbative matrix element $<0|\chi^\dagger \xi a^\dagger_Ha_H \xi^\dagger \chi |0>$
is evaluated at the origin.

However, unlike higher order pQCD calculation of color singlet S-wave heavy
quark-antiquark production/anihillation, the higher order pQCD calculation of color singlet P-wave heavy
quark-antiquark production/anihillation contains non-canceling infrared divergence due to real soft gluons
emission/absorption \cite{bar2,bar3,bar4}. Hence the non-perturbative effects can not be factored into the derivative of the non-relativistic wave function at the origin
 \cite{nrqcd}
\bea
|R'(0)|^2 = \frac{2\pi }{27}<0|{\cal O}_H|0>=\frac{2\pi }{27} <0|\chi^\dagger {\bar{\bf \nabla}} \xi a^\dagger_H \cdot a_H \xi^\dagger {\bar{\bf \nabla}} \chi |0>
\label{pw}
\eea
where
\bea
\chi^\dagger {\bar {\bf \nabla}} \xi = \chi^\dagger (\vec{\nabla} \xi)-(\vec{\nabla } \chi)^\dagger \xi
\label{lra}
\eea
and the non-perturbative matrix element $<0|\chi^\dagger {\bar{\bf \nabla}} \xi a^\dagger_H \cdot a_H \xi^\dagger {\bar{\bf \nabla}} \chi |0>$
is evaluated at the origin.

This is also easy to see from eq. (\ref{pw}) because, unlike $|R(0)|^2$ in eq. (\ref{sw}) which is
gauge invariant, the $|R'(0)|^2$ in eq. (\ref{pw}) is not gauge invariant. Since the
issue of gauge invariance and non-canceling infrared divergences due to real soft gluons emission/absorption
are related \cite{nayaksterman,ns1}, one expects that the gauge links need to be
supplied in eq. (\ref{pw}) to make $|R'(0)|^2$ gauge invariant and consistent
with the factorization of such non-canceling infrared divergences due to the
real soft gluons emission/absorption. In this paper we will investigate this in detail.

We find that the correct definition of the gauge invariant non-perturbative matrix element of the
color singlet P-wave heavy quarkonium production which is consistent with factorization of infrared divergences
is given by
\bea
<0|{\cal O}_H|0>= <0|\chi^\dagger \Phi {\bar {\bf \nabla}} \Phi^\dagger \xi a^\dagger_H \cdot a_H \xi^\dagger \Phi {\bar {\bf \nabla}} \Phi^\dagger \chi |0>
\label{pwc}
\eea
where
\bea
\Phi(x) ={\cal P}e^{-igT^c\int_0^\infty d\lambda l \cdot A^c(x+\lambda l)}
\label{glf}
\eea
is the gauge link in the fundamental representation of SU(3) and
\bea
\chi^\dagger \Phi {\bar {\bf \nabla}} \Phi^\dagger \xi  = \chi^\dagger \Phi (\vec{\nabla} \Phi^\dagger \xi)-(\vec{\nabla }\Phi^\dagger \chi)^\dagger \Phi^\dagger \xi.
\label{lrag}
\eea

We find that the long-distance behavior of the color singlet P-wave non-perturbative matrix element
$<0|\chi^\dagger \Phi {\bar {\bf \nabla}} \Phi^\dagger \xi a^\dagger_H \cdot a_H \xi^\dagger \Phi {\bar {\bf \nabla}} \Phi^\dagger \chi |0>$
in eq. (\ref{pwc}) is independent of the light-like vector $l^\mu$ at all order in coupling constant in QCD where $l^\mu$ is
used to define the gauge link in eq. (\ref{glf}).

Note that there is no gauge link in $|R(0)|^2$ in eq. (\ref{sw}) in case of S-wave heavy quarkonium because
there are no derivative operators like that in eq. (\ref{pwc}) for P-wave heavy quarkonium. The $|R(0)|^2$ in
eq. (\ref{sw}) is gauge invariant without gauge links which is consistent with the fact that there are no uncanceled
infrared divergences due to real soft gluons emission/absorption in case of color singlet S-wave heavy quarkonium.

In this paper we will provide a derivation of eq. (\ref{pwc}).

The paper is organized as follows. In section II we briefly review the appearance of infrared divergences
in higher order pQCD calculation in color singlet P-wave heavy quarkonium. In section III we discuss the
lack of gauge invariance in S-wave NRQCD color octet non-perturbative matrix element which is used to address
the issue of non-canceling infrared divergences in the color singlet P-wave heavy quarkonium production in the literature.
In section IV we describe the infrared divergence due to real soft gluon (eikonal gluon) emission/absorption in pQCD. In section V
we prove the factorization of infrared divergences in color singlet P-wave heavy quarkonium production at all orders in
coupling constant at high energy colliders. In section VI we derive the
correct definition of the gauge invariant non-perturbative matrix element of the color singlet P-wave heavy quarkonium production
as given by eq. (\ref{pwc}). Section VII contains conclusions.

\section{ Infrared Divergence in Higher Order pQCD Calculation in Color Singlet P-Wave Heavy Quarkonium }

In the higher order pQCD calculation of heavy quark-antiquark annihilation in the color singlet P-wave state,
the non-canceling infrared divergences were found for the color singlet P-wave heavy quarkonium decay
to hadrons \cite{bar2,bar3,bar4}. Similarly in the higher order pQCD calculation of hadroproduction of
heavy quark-antiquark in the color singlet P-wave state, the non-canceling infrared divergences were found
\cite{cac}. A satisfactory explanation of
these non-canceling infrared divergences in the case of color singlet P-wave heavy quarkonium production/decay
has not been given in the literature, partly because of the lack of understanding of the correct definition
of the gauge invariant non-perturbative matrix element of the color singlet P-wave heavy quarkonium which
is required to cancel these infrared divergences (factorization).

In addition to this it should be noted that the diagrammatic pQCD calculation is not always
sufficient to predict the correct definition of non-perturbative matrix element in QCD, no matter how many
orders of pQCD calculation is performed. On the other hand the non-perturbative QCD method (the path integral
formulation of QCD) can correctly predict the correct definition of the non-perturbative matrix element
in QCD. We will use path integral formulation of QCD in this paper.

In case of $h_c$ decays to hadrons the partonic process in which the non-canceling infrared divergence is found is
given by \cite{sch}
\bea
^1P_1 \rightarrow ggg.
\label{nia}
\eea
Similarly in case of $\chi_{cJ}$ decay to hadrons the partonic process in which the non-canceling infrared
divergence is found is given by \cite{bar2,bar3,bar4}
\bea
^3P_J \rightarrow q{\bar q}g.
\label{nid}
\eea
Note that the collinear divergences are canceled by the virtual corrections to the two gluon decay process \cite{cac}.
Hence the non-canceling infrared divergence is due to the emission/absorption of real soft gluon (eikonal gluon).

Similar to the hadronic decay of color singlet P-wave heavy quarkonium, the non-canceling infrared divergences
are found in the hadroproduction of color singlet P-wave heavy quarkonium. For example, the non-canceling
infrared divergence appears in the partonic process \cite{cac}
\bea
q{\bar q} \rightarrow ~^3P_J g.
\label{nip}
\eea
Note that unlike decay process in eq. (\ref{nid}) there is no collinear divergence in the production process in eq.
(\ref{nip}) and hence there is no virtual correction present either. The non-canceling infrared divergence
is due to emission/absorption of real soft gluon (eikonal gluon).

It can be noted that in the original calculation \cite{bar2,bar3,bar4} the binding energy was used as
infrared cutoff which can be mapped into the corresponding infrared divergence analysis in the dimensional
regularization scheme, see for example \cite{bc}.

Hence whether it is color singlet P-wave heavy quarkonium production or decay, the non-canceling infrared
divergences occur due to real soft gluons emission/absorption. In this paper we will derive the correct definition
of the gauge invariant color singlet P-wave non-perturbative matrix element which cancels
these infrared divergences at all order in coupling constant in QCD.

We emphasize here that we are not calculating the finite part of the cross section of
the color singlet P-wave heavy quarkonium production in the partonic level processes like that
in eq. (\ref{nip}). What we are investigating in this paper is to study the exact behavior of the infrared
divergences due to real soft gluons emission/absorption without modifying the finite value of the
cross section. Our main aim is to derive
the correct definition of the gauge invariant color singlet P-wave non-perturbative matrix element
which cancels these infrared divergences at all order in coupling constant.

The color singlet P-wave non-perturbative matrix element which was used in the earlier studies is given by
eq. (\ref{pw}) which is not gauge invariant. Hence it is no surprise that the non-canceling infrared divergences
were present in these earlier studies of color singlet P-wave heavy quarkonium production/decay.
It is expected that if the gauge links are supplied in the definition of the non-perturbative matrix
element to make it gauge invariant then it can accommodate these non-canceling infrared divergences \cite{nayaksterman,ns1}.
Hence when we use the gauge invariant definition of the color singlet P-wave non-perturbative matrix element as given
by eq. (\ref{pwc}) then we do not expect any uncanceled infrared divergences due to real soft gluons emission/absorption
in the color singlet P-wave heavy quarkonium production.

It is useful to mention here that the heavy quarkonium is also an useful signature \cite{qgp1} to study production
and detection of quark-gluon plasma at RHIC and LHC. In high energy heavy-ion collisions
at RHIC and LHC the study of heavy quarkonium production is more complicated than that in $pp$ collisions
due to the presence of QCD medium in non-equilibrium.

\section{ Lack of Gauge Invariance in NRQCD S-Wave Color Octet Non-Perturbative Matrix Element }

It can be mentioned here that in the NRQCD approach of heavy quarkonium study the S-wave color octet
non-perturbative matrix element
\bea
<0|{\cal O}_H|0>=<0|\chi^\dagger T^a \xi a^\dagger_H a_H \xi^\dagger T^a \chi |0>
\label{oc}
\eea
is used to factorize the above mentioned non-canceling infrared divergences which arise due to real soft gluons
emission/absorption (eikonal gluons) in the pQCD partonic level calculation in the color singlet P-wave heavy
quarkonium production \cite{nrqcd}. The NRQCD definition of the color singlet P-wave non-perturbative
matrix element used for this purpose is given by \cite{nrqcd}
\bea
<0|{\cal O}_H|0>=<0|\chi^\dagger {\bar{\bf D}} \xi a^\dagger_H \cdot a_H \xi^\dagger {\bar{\bf D}} \chi |0>
\label{oc1}
\eea
which is gauge invariant where ${\bf D}$ is the covariant derivative.

Note that in QCD the definition of the color singlet P-wave non-perturbative matrix element of the heavy quarkonium
is given by eq. (\ref{pw}) which contains the ordinary derivative ${\bf \nabla}$ whereas in NRQCD the definition of the color singlet P-wave
non-perturbative matrix element of the heavy quarkonium is given by eq. (\ref{oc1}) which contains covariant derivative
${\bf D}$. The main motivation to use covariant derivative ${\bf D}$ in NRQCD instead of ordinary derivative ${\bf \nabla}$ was to make the
the definition in eq. (\ref{oc1}) gauge invariant. However, it has to be remembered that it is the relative momentum ${\bf q}$
of the heavy quark-antiquark pair which appears in the definition of the color singlet P-wave non-prturbative matrix element
\cite{bar2,bar3,bar4} which can be obtained if ordinary derivative ${\bf \nabla}$ operates on the heavy quark field which can be
seen from the definition in eq. (\ref{pw}) in QCD.

Of course the appearance of ordinary derivative ${\bf \nabla}$ makes the definition of
the color singlet P-wave non-perturbative matrix element of heavy quarkonium in eq. (\ref{pw}) gauge non-invariant which is
the reason why we have non-canceling infrared divergences because the non-canceling infrared divergences due to real soft gluons
(eikonal gluons) emission/absorption can not be absorbed into the gauge non-invariant definition of the color singlet P-wave
non-perturbative matrix element of heavy quarkonium in eq. (\ref{pw}).
When gauge links are present in the gauge invariant definition of the color singlet
P-wave non-perturbative matrix element of heavy quarkonium as given by eq. (\ref{pwc}) then the non-canceling infrared divergences
due to the real soft gluons (eikonal gluons) emission/absorption can be absorbed into this gauge invariant definition of the color singlet
P-wave non-perturbative matrix element of heavy quarkonium in eq. (\ref{pwc}). Hence, unlike NRQCD which brings color octet
mechanism to solve infrared divergences problem in color singlet mechanism, we do not bring color octet mechanism to solve infrared divergences problem in color singlet mechanism.

In any case let us briefly discuss the issue of lack of gauge invariance of the S-wave color octet non-perturbative matrix
element of the heavy quarkonium production in NRQCD. In order for the non-canceling infrared divergences due to real soft
gluons (eikonal gluons) emission/absorption in the color singlet P-wave heavy
quarkonium production to be factorized into the definition of the NRQCD
S-wave color octet non-perturbative matrix element as given by eq. (\ref{oc}), the two non-perturbative matrix elements
in NRQCD in eqs. (\ref{oc}) and (\ref{oc1}) are related by the equation (see eq. (6.5) of \cite{nrqcd})
\bea
\Lambda \frac{d}{d\Lambda} <0|\chi^\dagger T^a \xi a^\dagger_H a_H \xi^\dagger T^a \chi |0> = \frac{4C_F \alpha_s(\Lambda)}{3N_c\pi M^2}<0|\chi^\dagger {\bar{\bf D}} \xi a^\dagger_H \cdot a_H \xi^\dagger {\bar{\bf D}} \chi |0>
\label{oc2}
\eea
which has the problem of gauge invariance in NRQCD because the left hand side of eq. (\ref{oc2}) is not
gauge invariant whereas the right hand side of eq. (\ref{oc2}) is gauge invariant where $\Lambda$ is the
ultraviolet cutoff of NRQCD. This is because the definition of the NRQCD S-wave color octet non-perturbative
matrix element as given by eq. (\ref{oc}) in NRQCD in \cite{nrqcd} is not gauge invariant. Note that the gauge links
are supplied in the NRQCD S-wave color octet non-perturbative matrix element in \cite{nayaksterman,ns1,nknr} to make it
gauge invariant and consistent with factorization of infrared divergences.

It can be mentioned here that in NRQCD, the (the long distance matrix elements) LDMEs are organized as powers of the relative velocity $(v)$ of the heavy quarks in the heavy-quarkonium states. At leading order in $v^2$, $|R'(0)|^2$ in eq. (\ref{pw}), which is usually calculated in potential models in Coulomb gauge, is gauge invariant and so is the evolution equation eq. (\ref{oc2}).

\section{ Infrared divergence due to real soft gluon (eikonal gluon) emission/absorption in QCD }

As mentioned in section II the non-canceling infrared divergences occur in the color singlet P-wave
heavy quarkonium production in the processes like that in eq. (\ref{nip}) due to real soft gluons
(eikonal gluons) emission/absorption. Hence in this section we will describe the infrared divergence due to real
soft gluon (eikonal) emission/absorption in QCD. Before describing infrared divergence due to soft gluon
(eikonal gluon) emission/absorption in QCD let us first describe the infrared divergence due to soft photon
(eikonal photon) emission/absorption in QED.

Consider the emission of a real photon with four momentum $k^\mu$ from an electron of four momentum $p^\mu$
in QED. The Feynman diagram contribution from this real photon emission process in QED is given by \cite{grammer}
\bea
\frac{1}{ {\not p} -{\not k} -m} {\not \epsilon}(k) u(p) = -\frac{ {\not p} -{\not k} +m}{2p\cdot k} {\not \epsilon}(k) u(p)
= -\frac{ p \cdot \epsilon(k)}{p\cdot k} u(p)+ \frac{{\not k} {\not \epsilon}(k)}{2p\cdot k}  u(p)
\label{gy}
\eea
where we have used
\bea
({\not p} -m)u(p) =0.
\eea
The photon field $\epsilon^\mu(k)$ can be written as
\bea
\epsilon^\mu(k) = [\epsilon^\mu(k) - \frac{k^\mu}{p \cdot k} p \cdot \epsilon(k)] +  \frac{k^\mu}{p \cdot k} p \cdot \epsilon(k) =\epsilon^\mu_{phys}(k)+\epsilon^\mu_{pure}(k)
\label{gy1}
\eea
where $\epsilon^\mu_{phys}(k)$ is the transversely polarized (physical) photon field and
$\epsilon^\mu_{pure}(k)=\frac{k^\mu}{p \cdot k} p \cdot \epsilon(k)$
is the (unphysical) longitudinally polarized pure gauge photon field.

Hence we expect that it is the transversely polarized (physical) photon field $\epsilon^\mu_{phys}(k)$
in eq. (\ref{gy1}) that contributes to the physical (finite) value of the cross section but
the (unphysical) longitudinally polarized pure gauge photon field $\epsilon^\mu_{pure}(k)$ does not contribute to
the physical (finite) value of the cross section. This can be seen by using eq. (\ref{gy1}) in (\ref{gy}) to find
\bea
\frac{{\not k} {\not \epsilon}(k)}{2p\cdot k}= \frac{{\not k} {\not \epsilon_{phys}}(k)}{2p\cdot k} \rightarrow {\rm finite}~~~~~~~~~{\rm when}~~~~~~~k^\mu \rightarrow 0
\label{gy3}
\eea
and
\bea
\frac{{\not k} {\not \epsilon_{pure}}(k)}{2p\cdot k} =0.
\label{gy3a}
\eea
Eq. (\ref{gy3}) is the non-eikonal contribution from this soft photon emission
Feynman diagram which contributes to the finite value of the physical cross section
whereas from eq. (\ref{gy3a}) we find that the (unphysical) longitudinally polarized
pure gauge photon field $\epsilon^\mu_{pure}(k)$ can not contribute to the finite value
of the physical cross section. The (unphysical) longitudinally polarized
pure gauge photon field $\epsilon^\mu_{pure}(k)$ in eq. (\ref{gy1}) accounts for the infrared divergence
from this soft photon emission Feynman diagram. This can be seen as follows.

By using eq. (\ref{gy1}) in (\ref{gy}) we find
\bea
\frac{ p \cdot \epsilon(k)}{p\cdot k} = \frac{ p \cdot \epsilon_{pure}(k)}{p\cdot k} \rightarrow \infty~~~~~~~~~~{\rm when}~~~~~~~~k^\mu \rightarrow 0.
\label{gy2}
\eea
and
\bea
\frac{ p \cdot \epsilon_{phys}(k)}{p\cdot k}=0.
\label{gy2a}
\eea
Eq. (\ref{gy2}) gives the infrared divergence in the eikonal approximation from this real soft
photon (eikonal photon) emission Feynman diagram. From eq. (\ref{gy2a}) we find that
the transversely polarized (physical) photon field $\epsilon^\mu_{phys}(k)$ in eq. (\ref{gy1}) does not contribute to
the infrared divergence from this real soft photon emission Feynman diagram in the eikonal approximation.

Hence from eqs. (\ref{gy2}), (\ref{gy3a}) and (\ref{gy}) we find that if the charge produces pure gauge field then
the study of infrared divergence due to soft photon (eikonal photon) emission from that charge can be simplified in
quantum field theory by using pure gauge.

In our calculation the momentum $P_1^\mu$ of the heavy quark can be different from the momentum $P_2^\mu$ of the heavy antiquark where
\begin{equation}
{\vec P}_1 -{\vec P}_2 = 2m {\vec v}.
\end{equation}
If the photon interacts with the electron and positron simultaneously then that photon is the virtual photon and we can consider the photon propagator $G_{\mu \nu}(k)$ in that situation instead of the polarization vector $\epsilon^\mu(k)$ of the photon which we considered in eq. (\ref{gy1}) for the real photon emission from the electron only.

Hence if the photon is attached simultaneously to the electron of momentum $p^\mu$ and to the positron of momentum $q^\mu$ then the photon propagator $G^{\mu \nu}(k)$ can be split into the physical (transverse) propagator $G^{\mu \nu}_{\rm phys}(k)$ and the pure gauge (longitudinal) propagator  $G^{\mu \nu}_{\rm pure}(k)$ as follows
\begin{equation}
G^{\mu \nu}(k)=\frac{g^{\mu \nu}}{k^2}=G^{\mu \nu}_{\rm phys}(k)+G^{\mu \nu}_{\rm pure}(k)
\label{fpr}
\end{equation}
which is equivalent to eq. (\ref{gy1}) of our paper where in the infrared (IR) limit (see eq. (2.1) of \cite{grammer})
\begin{equation}
G^{\mu \nu}_{\rm phys}(k)=\frac{1}{k^2}[g^{\mu \nu}-\frac{p \cdot q}{(p\cdot k)(q \cdot k)} k^\mu k^\nu]
\label{fpr1}
\end{equation}
and
\begin{equation}
G^{\mu \nu}_{\rm pure}(k)=\frac{p \cdot q}{(p\cdot k)(q \cdot k)} \frac{k^\mu k^\nu }{k^2}.
\label{fpr2}
\end{equation}

For the virtual photon interacting simultaneously with the electron of momentum $p^\mu$ and the positron of momentum $q^\mu$ the eikonal Feynman rule for the infrared (IR) divergences is given by
\begin{equation}
\frac{p^\mu}{p\cdot k}G_{\mu \nu}(k)\frac{q^\nu}{q \cdot k}
\label{fpr3}
\end{equation}
(see, for example, Eq. (29) of \cite{ns1}).

Hence from eqs. (\ref{fpr}), (\ref{fpr1}), (\ref{fpr2}) and (\ref{fpr3}) we find
\begin{equation}
\frac{p^\mu}{p\cdot k}G_{\mu \nu}^{\rm pure}(k)\frac{q^\nu}{q \cdot k}\rightarrow \infty, ~~~~~~{\rm as}~~~~~~~~~~k\rightarrow 0
\end{equation}
and
\begin{equation}
\frac{p^\mu}{p\cdot k}G_{\mu \nu}^{\rm phys}(k)\frac{q^\nu}{q \cdot k}=0.
\end{equation}
The above equation is equivalent to eq. (\ref{gy2a}) of our paper. Hence one finds that the pure gauge part is the source of the infrared divergence.

It is well known that the light-like charge produces pure gauge field at all the time-space points $x^\mu$
except at the positions perpendicular to the direction of the motion of the charge at the time of closest
approach in classical mechanics \cite{collinssterman,nayakj,nayake}. This is also true in quantum field theory
which can be shown as follows.

The generating functional for the photon field $Q^\mu(x)$ in the presence of external source $J^\mu(x)$ is given by
\bea
Z[J] = \int [dQ] e^{i\int d^4x[- \frac{1}{4} [\partial_\mu Q_\nu(x) -\partial_\nu Q_\mu(x)]  [\partial^\mu Q^\nu(x) -\partial^\nu Q^\mu(x)] -\frac{1}{2 \alpha} (\partial_\nu Q^{\nu })^2+J_\mu(x) Q^\mu(x)]}
\label{gf}
\eea
which gives the effective action \cite{nknr,peter}
\bea
<0|0>_J=\frac{Z[J]}{Z[0]} = e^{iS_{eff}},~~~~~~~~~~~~~~~~S_{eff} = -\frac{1}{2} \int d^4x J_\mu(x) \frac{1}{\partial^2} J^\mu(x) =\int d^4x {\cal L}_{eff}(x)\nonumber \\
\label{seff}
\eea
where ${\cal L}_{eff}(x)$ is the effective lagrangian density.

The eikonal current density $J^\mu_{eik}(x)$ of the charge with light-like four velocity $l^\mu$ can be obtained from the eikonal term of
eq. (\ref{gy})
\bea
e\int \frac{d^4k}{(2\pi)^4} \frac{ l \cdot Q(k)}{l\cdot k+i\epsilon }= -i \int d^4x J^\mu_{eik}(x) Q_\mu(x)
\label{gy4}
\eea
which gives the light-like eikonal current density
\bea
J^\mu_{eik}(x) = e\int_0^\infty d\lambda l^\mu \delta^{(4)}(x-\lambda l).
\label{ekc}
\eea
Using eq. (\ref{ekc}) in (\ref{seff}) we find
\bea
{\cal L}^{eik}_{eff}(x) = \frac{e^2}{2} [\frac{l^2}{(l\cdot x)^2}]^2=0,~~~~~~~~~~~~~~~l^2=0,~~~~~~~~~~~~~x \cdot l \neq 0
\label{leff1}
\eea
which means the eikonal current of light-like charge produces pure gauge field in quantum field theory
at all the time-space positions $x^\mu$ except at the positions perpendicular to the direction of the
motion of the charge at the time of closest approach ($l \cdot x \neq 0$) which agrees with the
corresponding result of the classical mechanics.

From eq. (\ref{gy}) the non-eikonal contribution
\bea
e\gamma^\mu \gamma^\nu \int \frac{d^4k}{(2\pi)^4} \frac{k_\mu Q_\nu(k)}{2p\cdot k+i\epsilon }=  \int d^4x J^\mu_{noneik}(x) Q_\mu(x)
\label{nek}
\eea
gives the non-eikonal current density
\bea
J^\mu_{noneik}(x) = e\gamma^\nu \gamma^\mu \int_0^\infty d\lambda \partial_\nu \delta^{(4)}(x-\lambda p).
\label{nekc}
\eea
Using eqs. (\ref{nekc}) and (\ref{ekc}) in eq. (\ref{seff}) we find that the effective interaction lagrangian density
due to the interaction between the (light-like or non-light-like)
non-eikonal current and the gauge field generated by the light-like eikonal current is given by
\bea
{\cal L}^{int}_{eff}(x) =l^2\frac{e^2}{2} \frac{(p\cdot l) (p \cdot x) -p^2 l \cdot x}{(l \cdot x)^3~ [(p \cdot x)^2 - p^2x^2]^{\frac{3}{2}}}=0,~~~~~~~l^2=0,~~~~~~~x \cdot l \neq 0,~~~~~~~p\cdot x \neq 0\nonumber \\
\label{ileff1}
\eea
which implies that the light-like eikonal line can be replaced by the pure gauge background field to study factorization
of infrared divergences in quantum field theory without modifying the finite value of the physical cross section.

Hence we find that since the light-like eikonal
current produces pure gauge field, the study of infrared divergence due to soft photon (eikonal photon) emission
due to the presence of light-like eikonal line can be simplified in quantum field theory by using pure gauge
without modifying the finite value of the cross section.
Since the light-like eikonal current produces pure gauge field in classical mechanics and in quantum field theory,
the factorization of infrared divergence due to the presence of light-like eikonal line can be studied by using
path integral formulation of the background field method of quantum field theory
in the presence of pure gauge background field \cite{nayakarx,nknr,n3,tucci}.

In QED the light-like electric charge produces U(1) pure gauge field $A^\mu(x)$ given by
\bea
A_\mu(x)=\partial_\mu \omega(x).
\eea
In QCD the light-like color charges produce SU(3) pure gauge field given by \cite{nayakarx,nknr,n3}
\bea
T^cA^{\mu c}(x)= \frac{1}{ig} [\partial^\mu \Phi(x)]\Phi^{-1}(x)
\label{pug}
\eea
where $\Phi(x)$ is the gauge link given by eq. (\ref{glf}) where $l^\mu$ is the light-like four velocity.

\section{ Proof of factorization of Infrared Divergence in Color Singlet P-Wave heavy quarkonium production }

As discussed above the factorization of infrared divergences due to the soft gluons emission/absorption with the
light-like eikonal line can be studied by using the path integral formulation of the background field method of
QCD \cite{nayakarx,nknr,n3} in the presence of SU(3) pure gauge background field $A^{\mu a}(x)$ as given by eq. (\ref{pug}).
We follow the notation of \cite{thooft,zuber,abbott} and denote the quantum gluon field by $Q^{\mu a}$ and the
background field by $A^{\mu a}$.

In the path integral formulation of QCD we find \cite{muta,abbott}
\bea
&&<0|{\bar \Psi}(x') {\bar{\bf \nabla}}_{x'} \Psi(x') \cdot {\bar \Psi}(x'') {\bar{\bf \nabla}}_{x''} \Psi(x'')|0>=\int \Pi_{l=1}^3 [d{\bar \psi}_l] [d \psi_l ] [d{\bar \Psi}] [d \Psi ][dQ] \nonumber \\
&&{\bar \Psi}(x') {\bar{\bf \nabla}}_{x'} \Psi(x') \cdot {\bar \Psi}(x'') {\bar{\bf \nabla}}_{x''} \Psi(x'') ~{\rm det}(\frac{\delta \partial_\mu Q^{\mu c}}{\delta \omega^d}) \nonumber \\
&&e^{i\int d^4x [\sum_{l=1}^3 {\bar \psi}_l [i\gamma^\mu \partial_\mu -m_l +gT^a\gamma^\mu Q^a_\mu] \psi_l +{\bar \Psi} [i\gamma^\mu \partial_\mu -M +gT^a\gamma^\mu Q^a_\mu] \Psi-\frac{1}{4}{F^a}_{\mu \nu}^2[Q] -\frac{1}{2 \alpha} (\partial_\mu Q^{\mu a})^2 ]}
\label{avg1}
\eea
and in the path integral formulation of the background field method of QCD we find \cite{thooft,abbott,zuber}
\bea
&& <0|{\bar \psi}(x') {\bar{\bf \nabla}}_{x'} \psi(x') \cdot {\bar \psi}(x'') {\bar{\bf \nabla}}_{x''} \psi(x'')|0>_A=\int \Pi_{l=1}^3 [d{\bar \psi}_l] [d \psi_l ] [d{\bar \Psi}] [d \Psi ][dQ]\nonumber \\
&&{\bar \psi}(x') {\bar{\bf \nabla}}_{x'} \psi(x') \cdot {\bar \psi}(x'') {\bar{\bf \nabla}}_{x''} \psi(x'')]~{\rm det}(\frac{\delta G^c(Q)}{\delta \omega^d}) \nonumber \\
&&e^{i\int d^4x [\sum_{l=1}^3 {\bar \psi}_l [i\gamma^\mu \partial_\mu -m_l +gT^a\gamma^\mu (A+Q)^a_\mu] \psi_l
+{\bar \Psi} [i\gamma^\mu \partial_\mu -M +gT^a\gamma^\mu (A+Q)^a_\mu] \Psi -\frac{1}{4}{F^a}_{\mu \nu}^2[A+Q] -\frac{1}{2 \alpha}
(G^a(Q))^2]}\nonumber \\
\label{avg2}
\eea
where
\bea
F_{\mu \nu}^c[A+Q]=\partial_\mu [A_\nu^c+Q_\nu^c]-\partial_\nu [A_\mu^c+Q_\mu^c]+gf^{cba} [A_\mu^b+Q_\mu^b][A_\nu^a+Q_\nu^a],~~~~~~F^2_{\mu \nu}=F_{\mu \nu}F^{\mu \nu}, \nonumber \\
\eea
and
\bea
G^c(Q) =\partial_\mu Q^{\mu c} + gf^{cba} A_\mu^b Q^{\mu a}
\label{gab}
\eea
is the gauge fixing term and the type I gauge transformation in the background field method of QCD is given by
\bea
&& T^c A'^{\mu c} = \Phi T^cA^{\mu c}\Phi^{-1} +\frac{1}{ig} (\partial^\mu \Phi)\Phi^{-1} \nonumber \\
&& T^c Q'^{\mu c} = \Phi T^cQ^{\mu c}\Phi^{-1}.
\label{gt1}
\eea
In eqs. (\ref{avg1}) and (\ref{avg2}) the $\psi_l$ is the Dirac field of the light quark of flavor $l=1,2,3=u,d,s$ of mass $m_l$, the $\Psi$ is the
Dirac field of the heavy quark of mass $M$ and ${\bar{\bf \nabla}}$ is defined by eq. (\ref{lra}).

Changing the integration variable $Q \rightarrow Q-A$ inside the path integration in eq. (\ref{avg2}) we find
\bea
&& <0|{\bar \psi}(x') {\bar{\bf \nabla}}_{x'} \psi(x') \cdot {\bar \psi}(x'') {\bar{\bf \nabla}}_{x''} \psi(x'')|0>_A=\int \Pi_{l=1}^3 [d{\bar \psi}_l] [d \psi_l ] [d{\bar \Psi}] [d \Psi ][dQ]\nonumber \\
&&{\bar \psi}(x') {\bar{\bf \nabla}}_{x'} \psi(x') \cdot {\bar \psi}(x'') {\bar{\bf \nabla}}_{x''} \psi(x'')]~{\rm det}(\frac{\delta G^c_f(Q)}{\delta \omega^d}) \nonumber \\
&&e^{i\int d^4x [\sum_{l=1}^3 {\bar \psi}_l [i\gamma^\mu \partial_\mu -m_l +gT^a\gamma^\mu Q^a_\mu] \psi_l
+{\bar \Psi} [i\gamma^\mu \partial_\mu -M +gT^a\gamma^\mu Q^a_\mu] \Psi -\frac{1}{4}{F^a}_{\mu \nu}^2[Q] -\frac{1}{2 \alpha}
(G^a_f(Q))^2]}
\label{avg3}
\eea
where
\bea
G^c_f(Q) =\partial_\mu Q^{\mu c} + gf^{cba} A_\mu^b Q^{\mu a}-\partial_\mu A^{\mu c}
\label{gfab}
\eea
and eq. (\ref{gt1}) becomes
\bea
&& T^c Q'^{\mu c} = \Phi T^cQ^{\mu c}\Phi^{-1} +\frac{1}{ig} (\partial^\mu \Phi)\Phi^{-1}.
\label{gt2}
\eea
Since the change of integration variables from unprimed variables to primed variables does not change the value of the integration we
find from eq. (\ref{avg3})
\bea
&& <0|{\bar \psi}(x') {\bar{\bf \nabla}}_{x'} \psi(x') \cdot {\bar \psi}(x'') {\bar{\bf \nabla}}_{x''} \psi(x'')|0>_A=\int \Pi_{l=1}^3 [d{\bar \psi}'_l] [d \psi'_l ] [d{\bar \Psi}'] [d \Psi' ][dQ']\nonumber \\
&&{\bar \psi}'(x') {\bar{\bf \nabla}}_{x'} \psi'(x') \cdot {\bar \psi}'(x'') {\bar{\bf \nabla}}_{x''} \psi'(x'')]~{\rm det}(\frac{\delta G^c(Q')}{\delta \omega^d}) \nonumber \\
&&e^{i\int d^4x [\sum_{l=1}^3 {\bar \psi}'_l [i\gamma^\mu \partial_\mu -m_l +gT^a\gamma^\mu Q'^a_\mu] \psi'_l
+{\bar \Psi}' [i\gamma^\mu \partial_\mu -M +gT^a\gamma^\mu Q'^a_\mu] \Psi' -\frac{1}{4}{F^a}_{\mu \nu}^2[Q'] -\frac{1}{2 \alpha}
(G^a_f(Q'))^2]}.
\label{avg4}
\eea
From
\bea
\psi' = \Phi \psi
\label{gt3}
\eea
and from eq. (\ref{gt2}) we find \cite{nayakarx,nknr,n3}
\bea
&& [dQ'] =[dQ],~~~~~~~~~~~[d{\bar \psi}'_{l}] [d \psi'_{l} ]=[d{\bar \psi}_{l}] [d \psi_{l} ],~~~~~~~~~~~~[d{\bar \Psi}'] [d \Psi' ]=[d{\bar \Psi}] [d \Psi ],\nonumber \\
&&{\bar \psi}'_{l} [i\gamma^\mu \partial_\mu -m_l +gT^c\gamma^\mu Q'^c_{\mu }] \psi'_{l}={\bar \psi}_{l} [i\gamma^\mu \partial_\mu -m_l +gT^c\gamma^\mu Q^c_{\mu }] \psi_{l}, \nonumber \\
&&{\bar \Psi}' [i\gamma^\mu \partial_\mu -M +gT^c\gamma^\mu Q'^c_{\mu}] \Psi'={\bar \Psi} [i\gamma^\mu \partial_\mu -M +gT^c\gamma^\mu Q^c_{\mu}]\Psi,~~~~~~~~~{F}^2[Q']={F}^2[Q] \nonumber \\
&& (G_f^c(Q'))^2 = (\partial_\mu Q^{\mu c})^2,~~~~~~~~~~~{\rm det} [\frac{\delta G_f^c(Q')}{\delta \omega^d}] ={\rm det}[\frac{ \delta (\partial_\mu Q^{\mu c})}{\delta \omega^d}].
\label{gt4}
\eea

From eqs. (\ref{gt4}), (\ref{gt3}), (\ref{avg4}) and (\ref{avg1}) we find
\bea
&&<0|{\bar \Psi}(x') {\bar{\bf \nabla}}_{x'} \Psi(x') a^\dagger_H \cdot a_H {\bar \Psi}(x) {\bar{\bf \nabla}}_x \Psi(x)|0>\nonumber \\
&&=   <0|{\bar \Psi}(x') \Phi(x'){\bar{\bf \nabla}}_{x'} \Phi^\dagger(x') \Psi(x') a^\dagger_H \cdot a_H {\bar \Psi}(x) \Phi(x) {\bar{\bf \nabla}}_x \Phi^\dagger(x) \Psi(x)|0>_A
\label{fact}
\eea
which proves factorization of infrared divergences in color singlet P-wave heavy quarkonium production at all order in coupling constant in
QCD where the gauge link $\Phi(x)$ is given by eq. (\ref{glf}). Note that the ${\bar{\bf \nabla}}$ in the left hand side of eq.
(\ref{fact}) is defined by eq. (\ref{lra}) and the ${\bar{\bf \nabla}}$ in the right hand side of eq. (\ref{fact}) is defined by eq. (\ref{lrag}).

\section{ Correct definition of non-perturbative matrix element of color singlet P-wave heavy quarkonium Production }

From eq. (\ref{fact}) we find that the correct definition of the gauge invariant color singlet P-wave non-perturbative matrix element of
heavy quarkonium production which is consistent with factorization of infrared divergences in QCD at all orders in coupling constant is
given by
\bea
<0|{\cal O}_H|0> = <0|\chi^\dagger \Phi {\bar {\bf \nabla}} \Phi^\dagger \xi a^\dagger_H \cdot a_H \xi^\dagger \Phi {\bar {\bf \nabla}} \Phi^\dagger \chi |0>
\label{pwcf}
\eea
which reproduces eq. (\ref{pwc}) where the gauge link $\Phi(x)$ is given by eq. (\ref{glf}) and ${\bar{\bf \nabla}}$ is defined in eq. (\ref{lrag}).

Since the left hand side of eq. (\ref{fact}) is independent of the light-like vector $l^\mu$ one finds that the right hand side
of eq. (\ref{fact}) is independent of the light-like vector $l^\mu$ used to define the gauge link $\Phi(x)$ in eq. (\ref{glf}).
This proves that the long-distance behavior of the color singlet P-wave non-perturbative matrix element
$<0|\chi^\dagger \Phi {\bar {\bf \nabla}} \Phi^\dagger \xi a^\dagger_H \cdot a_H \xi^\dagger \Phi {\bar {\bf \nabla}} \Phi^\dagger \chi |0>$
in eq. (\ref{pwcf}) is independent of the light-like vector $l^\mu$ at all order in coupling constant in QCD.

This completes the derivation of the correct definition of the non-perturbative matrix element of color singlet
P-wave heavy quarkonium production which is gauge invariant and is consistent with the factorization of infrared
divergences at all order in coupling constant in QCD.

Note that the relationship between the color singlet P-wave long distance matrix element (LDME) and the derivative of the non-relativistic wave function at origin $|R'(0)|^2$ is very useful because it helps us to reduce the number of non-perturbative input parameters to be determined. Hence it is important to discuss how this relationship is affected by the redefinition in eq. (\ref{pwc}). First of all it is found that if one uses the derivative of the non-relativistic wave function at the origin $|R'(0)|^2$ in eq. (\ref{cssip}) then the infrared (IR) divergences do not cancel  \cite{cac,bar2,bar3,bar4}, {\it i. e.}, the non-perturbative effects can not be factored into the derivative of the non-relativistic wave function at the origin $|R'(0)|^2$. In order to cancel the infrared (IR) divergences and to be consistent with the factorization theorem at all orders in coupling constant the definition of the color singlet P-wave non-perturbative matrix element from eq. (\ref{pwc}) should be used in eq. (\ref{cssip}). Note that since the color singlet P-wave non-perturbative matrix element in eq. (\ref{pwc}) is a non-perturbative quantity it can not be studied by using perturbative QCD. Due to these infrared (IR) divergences, factorization and non-perturbative QCD issues there is no simple relation between the color singlet P-wave non-perturbative matrix element in eq. (\ref{pwc}) at all orders in coupling constant and the derivative of the non-relativistic wave function at the origin $|R'(0)|^2$. The color singlet P-wave non-perturbative matrix element in eq. (\ref{pwc}) at all orders in coupling constant can be studied by using the non-perturbative QCD. In the phenomenological studies of the infrared sensitive processes in color singlet P-wave heavy quarkonium production, the definition from eq. (\ref{pwc}) should be used in eq. (\ref{cssip}) to extract the non-perturbative matrix element from the experiments in order to be consistent with the factorization of the infrared (IR) divergences at all orders in coupling constant.

\section{Conclusions}
Recently we have proved factorization of infrared divergences in NRQCD S-wave
heavy quarkonium production at high energy colliders at all orders in coupling constant.
One of the problem which still exists in the higher order pQCD calculation of color singlet P-wave heavy
quarkonium production/anihillation is the appearance of non-canceling infrared divergences due to real soft gluons
exchange, although no such infrared divergences are present in the color singlet S-wave heavy quarkonium.

In this paper
we have found that since the non-perturbative matrix element of the color singlet P-wave
heavy quarkonium production contains derivative operators, the gauge links are necessary to make it gauge invariant and
be consistent with the factorization of such non-canceling infrared divergences at all orders in coupling constant.
In case of color singlet S-wave heavy
quarkonium production the gauge links cancel because of the absence of derivative operators which is consistent with
the absence of such non-canceling infrared divergences at all orders in coupling constant.


\begin{thebibliography}{99}
%
\bibitem{nknr} G. C. Nayak, Eur. Phys. J. C76 (2016) 448.

\bibitem{jpsi} J.J. Aubert {\it et al.}, Phys. Rev. Lett. 33 (1974) 1404; J.E. Augustin {\it et al.},
Phys. Rev. Lett. 33 (1974) 1406, Adv. Exp. Phys. 5 (1976) 141.

\bibitem{pqcd} T. Appelquist and H. D. Politzer, Phys. Rev. Lett. 34 (1975) 43; Phys. Rev. D12 (1975) 1404.

\bibitem{pqcd1} D. J. Gross and F. Wilczek, Phys. Rev. Lett. 30 (1973) 1343; H. D. Politzer, Phys. Rev. Lett. 30 (1973) 1346.

\bibitem{cac} A. Petrelli {et al}., Nucl. Phys. B514 (1998) 245, hep-ph/9707223.

\bibitem{bar1} R. Barbieri, {\it et al.}, Nucl. Phys. B154 (1979) 535.

\bibitem{bar2} R. Barbieri, {\it et al.}, Phys. Lett. 61B (1976) 465; R. Barbieri, {\it et al.}, Nucl. Phys. B162 (1980) 220.

\bibitem{bar3} R. Barbieri, {\it et al.}, Phys. Lett. 95B (1980) 93; R. Barbieri, {\it et al.}, Nucl. Phys. B192 (1981) 61.

\bibitem{bar4} W. Kwong, {\it et al.}, Phys. Rev. D37 (1988) 3210.

\bibitem{harris} I. Harris and L. M. Brown, Phys. Rev. 105 (1957) 1656.

\bibitem{lep} P. Mackenzie and P. Lepage, Phys. Rev. Lett. 47 (1981) 1244.

\bibitem{nrqcd} G. T. Bodwin, E. Braaten and G. Lepage, Phys. Rev. D51 (1995) 1125.

\bibitem{nayaksterman} G. C. Nayak, J. Qiu and G. Sterman, Phys. Lett. B613 (2005) 45; Phys. Rev. D72 (2005) 114012.

\bibitem{ns1} G. C. Nayak, J. Qiu and G. Sterman, hep-ph/0608066, Phys. Rev. D74 (2006) 074007.

\bibitem{sch} G. A. Schuler, CERN-TH.7170 (1994), hep-ph/9403387.

\bibitem{bc} E. Braaten and Y-Q Chen, Phys. Rev. D55 (1997) 7152, hep-ph/9701242.

\bibitem{qgp1} T. Matsui and H. Satz, Phys. Lett. B178 (1986) 416.

\bibitem{grammer} G. Grammer and D. R. Yennie, Phys. Rev. D8 (1973) 4332.

\bibitem{collinssterman} J. C. Collins, D. E. Soper and G. Sterman, Nucl. Phys. B261 (1985) 104.

\bibitem{nayakj} G. C. Nayak, JHEP1303 (2013) 001.

\bibitem{nayake} G. C. Nayak, Eur. Phys. J. C73 (2013) 2442.

\bibitem{peter}  G. C. Nayak and P. van Nieuwenhuizen, Phys. Rev. D71 (2005) 125001; G. C. Nayak, Phys. Rev.
D72 (2005) 125010; F. Cooper and G. C. Nayak, Phys. Rev. D73 (2006) 065005.

\bibitem{nayakarx} G. C. Nayak, Phys. Part. Nucl. Lett. 13 (2016) 417.

\bibitem{n3} G. C. Nayak, Phys. Part. Nucl. Lett. 14 (2017) 18.

\bibitem{tucci} R. Tucci, Phys. Rev. D32 (1985) 945.

\bibitem{abbott} L. F. Abbott, Nucl. Phys. B185 (1981) 189.

\bibitem{thooft} G. 't Hooft, Nucl. Phys. B62 (1973) 444.

\bibitem{zuber} H. Klueberg-Stern and J. B. Zuber, Phys. Rev. D12 (1975) 482.

\bibitem{muta} See for example, T. Muta, {\it Foundations of Quantum Chromodynamics}, World Scientific lecture notes in physics-Vol. 5.

\end{thebibliography}
\end{document}